\documentclass[12pt, conference, compsocconf, onecolumn, draftcls]{IEEEtran}
\usepackage[section] {placeins}
\usepackage{graphicx}
\usepackage{amsmath}
\usepackage{algorithm}
\usepackage{algorithmic}
\usepackage{subfigure}
\usepackage{rotating}
\usepackage{multirow}
\usepackage{array}
\usepackage{rotating}
\usepackage{auto-pst-pdf}
\usepackage{graphicx}
\usepackage{color}

\begin{document}
\title{On Adaptive Energy Efficient\\ Transmission in WSNs}

\author{M. Tahir{$^\S$}, N. Javaid{$^\S$}, A. Iqbal{$^\S$}, Z. A. Khan$^\ddag$, N. Alrajeh$^{\sharp}$\\
{$^\S$}COMSATS Institute of Information Technology, Islamabad, Pakistan\\
$^\ddag$Faculty of Engineering, Dalhousie University, Halifax, Canada.\\
$^{\sharp}$B.M.T, C.A.M.S, King Saud University, Riyadh, Saudi Arabia.\\}

\maketitle
\IEEEpeerreviewmaketitle

\begin{abstract}
One of the major challenges in design of Wireless Sensor Networks (WSNs) is to reduce energy consumption of sensor nodes to prolong lifetime of finite-capacity batteries. In this paper, we propose Energy-efficient Adaptive Scheme for Transmission (EAST) in WSNs. EAST is an IEEE 802.15.4 standard compliant. In this scheme, open-loop is used for temperature-aware link quality estimation and compensation. Whereas, closed-loop feedback process helps to divide network into three logical regions to minimize overhead of control packets. Threshold on transmitter power loss $(RSSI_{loss})$ and current number of nodes ($n_{c}(t)$) in each region help to adapt transmit power level ($P_{level}$) according to link quality changes due to temperature variation. Evaluation of propose scheme is done by considering  mobile sensor nodes and reference node both static and mobile. Simulation results show that propose scheme effectively adapts transmission $P_{level}$ to changing link quality with less control packets overhead and energy consumption as compared to classical approach with single region in which maximum transmitter $P_{level}$ assigned to compensate temperature variation.
\end{abstract}


\section{Introduction}

WSNs are currently being considered for many applications; including industrial, security surveillance, medical, environmental and weather monitoring. Due to limited battery lifetime at each sensor node; minimizing transmitter $P_{level}$ to increase energy efficiency and network lifetime is useful. Sensor nodes consist of three parts; sensing unit, processing unit and transceiver~\cite{1}. Limited battery requires low power sensing, processing and communication system. Energy efficiency is of paramount interest and optimal WSN  should consume minimum amount of power.

In WSNs, sensor nodes are widely deployed in different environments to collect data. As sensor nodes usually operate on limited battery, so each sensor node communicate using a low power wireless link and link quality varies significantly due to environmental dynamics like temperature, humidity etc. Therefore, while maintaining good link quality between sensor nodes we need to reduce energy consumption for data transmission to extend network lifetime~\cite{2}, ~\cite{3}, ~\cite{4}. IEEE802.15.4 is a standard used for low energy, low data rate applications like WSN. This standard operate at  frequency  2.45 GHz with  channels up to 16 and  data rate  250 kbps.

To efficiently compensate link quality changes due to temperature variations, we propose a new scheme for $P_{level}$ control EAST, that improves network lifetime while achieving required reliability between sensor nodes. This scheme is based on combination of open-loop and closed-loop feedback processes in which we divide network into three regions on basis of threshold on $RSSI_{loss}$ for each region. In open-loop process, each node estimates link quality using its temperature sensor. Estimated link quality degradation is then effectively compensated using closed-loop feedback process by applying propose scheme. In closed-loop feedback process, appropriate transmission $P_{level}$ control is obtained  which assign substantially less power than those required in existing transmission power control schemes.

Rest of the paper is organized as follows: section II briefs the related existing work and motivation for this work. In section III, we provide the readers with our proposed scheme. In section IV, we model our proposed scheme. Experimental results have been given in section V.

\section{Related Work and Motivation}

To transmit data efficiently over wireless channels in WSNs, existing schemes set some minimum transmission $P_{level}$ for maintaining reliability. These schemes either decrease interference among sensor nodes or increase unnecessary energy consumption. In order to adjust transmission $P_{level}$, reference node periodically broadcasts a beacon message. When nodes hear a beacon message from a reference node, nodes transmit an ACK message. Through this interaction, reference node estimate connectivity between nodes.

In Local Mean Algorithm (LMA), a reference node broadcasts LifeMsg message. Nodes transmit LifeAckMsg after they receive LifeMsg. Reference nodes count number of LifeAckMsgs and transmission $P_{level}$ to maintain appropriate connectivity. For example, if number of LifeAckMsgs is less than NodeMinThresh; transmission $P_{level}$ is increased. In contrast, if number of LifeAckMsgs is more than NodeMaxThresh transmission; $P_{level}$ is decreased. As a result, they provide improvement of network lifetime in a sufficiently connected network. However, LMA only guarantees connectivity between nodes and cannot estimate link quality~\cite{5}.

Local Information No Topology/Local Information Link-state Topology (LINT/LILT) and Dynamic Transmission Power Control (DTPC) use $RSSI_{loss}$ to estimate transmitter $P_{level}$. Nodes exceeding threshold $RSSI_{loss}$  are regarded as neighbor nodes with reliable links. Transmission $P_{level}$ also controlled by Packet Reception Ratio (PRR) metric. As for the neighbor selection method, three different methods have been used in the literature: connectivity based, PRR based and $RSSI_{loss}$ based. In LINT/LILT, a node maintains a list of neighbors whose $RSSI_{loss}$ values are higher than the threshold $RSSI_{loss}$, and it adjusts the radio transmission $P_{level}$ if number of neighbors is outside the predetermined bound. In LMA/LMN, a node determines its range by counting how many other nodes acknowledged to the beacon message it has sent ~\cite{6}.

Adaptive Transmission Power Control (ATPC) adjusts transmission $P_{level}$ dynamically according to spatial and temporal effects. This scheme tries to adapt link quality that changes over time by using closed-loop feedback. However, in large-scale WSNs, it is difficult to support scalability due to serious overhead required to adjust transmission $P_{level}$ of each link. The result of applying ATPC is that every node knows the proper transmission $P_{level}$ to use for each of its neighbors, and every node maintains good link qualities with its neighbors by dynamically adjusting the transmission $P_{level}$ through on-demand feedback packets. Uniquely, ATPC adopts a feedback-based and pairwise transmission $P_{level}$ control. By collecting the link quality history, ATPC builds a model for each neighbor of the node. This model represents an in-situ correlation between transmission $P_{level}$ and link qualities. With such a model, ATPC tunes the transmission $P_{level}$ according to monitored link quality changes. The changes of transmission $P_{level}$ reflect changes in the surrounding environment ~\cite{7}.

Existing approaches estimate variety of link quality indicators by periodically broadcasting a beacon message. In addition, feedback process is repeated for adaptively controlling transmission $P_{level}$. In adapting link quality for environmental changes, where temperature variation occur, packet overhead for transmission $P_{level}$ control should be minimized. Reducing number of control packets while maintaining reliability is an important technical issue ~\cite{8}.

Radio communication quality between low power sensor devices is affected by spatial and temporal factors. The spatial factors include the surrounding environment, such as terrain and the distance between the transmitter and the receiver. Temporal factors include surrounding environmental changes in general, such as weather conditions (Temperature). To establish an effective transmission $P_{level}$ control mechanism, we need to understand the dynamics between link quality and $RSSI_{loss}$ values. Wireless link quality refers to the radio channel communication performance between a pair of nodes. PRR is the most direct metric for link quality. However, the PRR value can only be obtained statistically over a long period of time. $RSSI_{loss}$ can be used effectively as binary link quality metrics for transmission $P_{level}$ control~\cite{9}.

Radio irregularity results in radio signal strength variation in different directions, but the signal strength at any point within the radio transmission range has a detectable correlation with transmission power in a short time period.  There are three main reasons for the fluctuation in the $RSSI_{loss}$. First, fading causes signal strength variation at any specific distance. Second, the background noise impairs the channel quality seriously when the radio signal is not significantly stronger than the noise signal. Third, the radio hardware doesn’t provide strictly stable functionality~\cite{10}.

Since the variation is small, this relation can be approximated by a linear curve. The correlation between $RSSI_{loss}$ and transmission $P_{level}$ is approximately linear.  Correlation between transmission $P_{level}$ and $RSSI_{loss}$ is largely influenced by environments, and this correlation changes over time. Both the shape and the degree of variation depend on the environment. This correlation also dynamically fluctuates when the surrounding environmental conditions change. The fluctuation is continuous, and the changing speed depends on many factors, among which the degree of environmental variation is one of the main factors~\cite{11}.

Propose energy efficient transmission scheme EAST helps efficiently compensate link quality changes due to temperature variation. To reduce packet overhead for adaptive power control temperature measured by sensors is utilized to adjust transmission $P_{level}$ for all  three regions based on $RSSI_{loss}$. Compared to single region in which large control packets overhead occur even due to small change in link quality. Closed-loop feedback process is executed to minimize control packets overhead and required transmitter $P_{level}$.

\section{Proposed Energy Efficient Transmission Scheme}

In this section, we present energy efficient transmission scheme that maintains link quality during temperature variation in wireless environment. It utilizes open-loop process based on sensed temperature information according to temperature variation. Closed-loop feedback process based on control packets is further used to accurately adjust transmission $P_{level}$. By adopting both open-loop and closed-loop feedback processes we divide network into three regions A, B, C for high, medium and low $RSSI_{loss}$ respectively.

%fig1
 \begin{figure}[h]
\begin{center}
\includegraphics[scale=0.9]{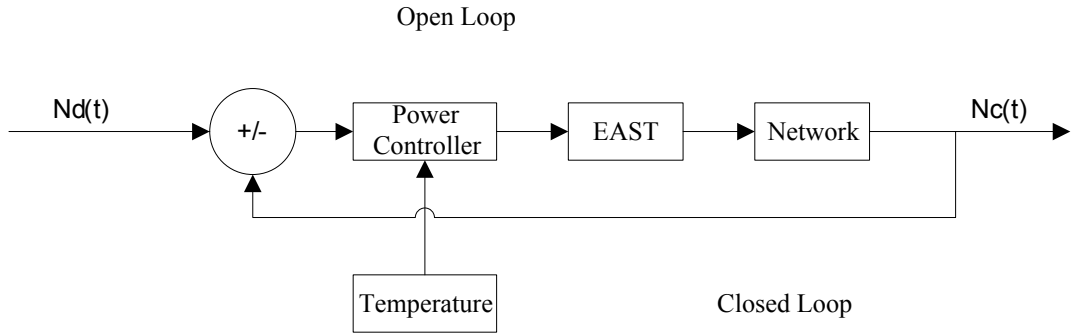}
\caption{Block Diagram}
\end{center}
\end{figure}

In order to assign minimum and reachable transmission $P_{level}$ to each link EAST is designed. EAST has two phases that is initial and run-time. In initial phase reference node build a model for nodes in network. In run-time phase based on previous model EAST adapt the link quality to dynamically maintain each link with respect to time. In a relatively stable network, control overhead occurs only in measuring link quality in initial phase. But in a relatively unstable network because link quality is continuously changing initial phase is repeated and serious overhead occur. Before we present block diagram for proposed scheme some variables are defined as follows (1)Current nodes in a region $n_{c}(t)$ (2) Desired nodes in a region $n_{d}(t)$ (3) Error: e(t) = $n_{d}(t) - n_{c}(t)$,(4) $P_{level}$.

%fig2
 \begin{figure*}[t]
\begin{center}
\includegraphics[height=18cm,width=15cm]{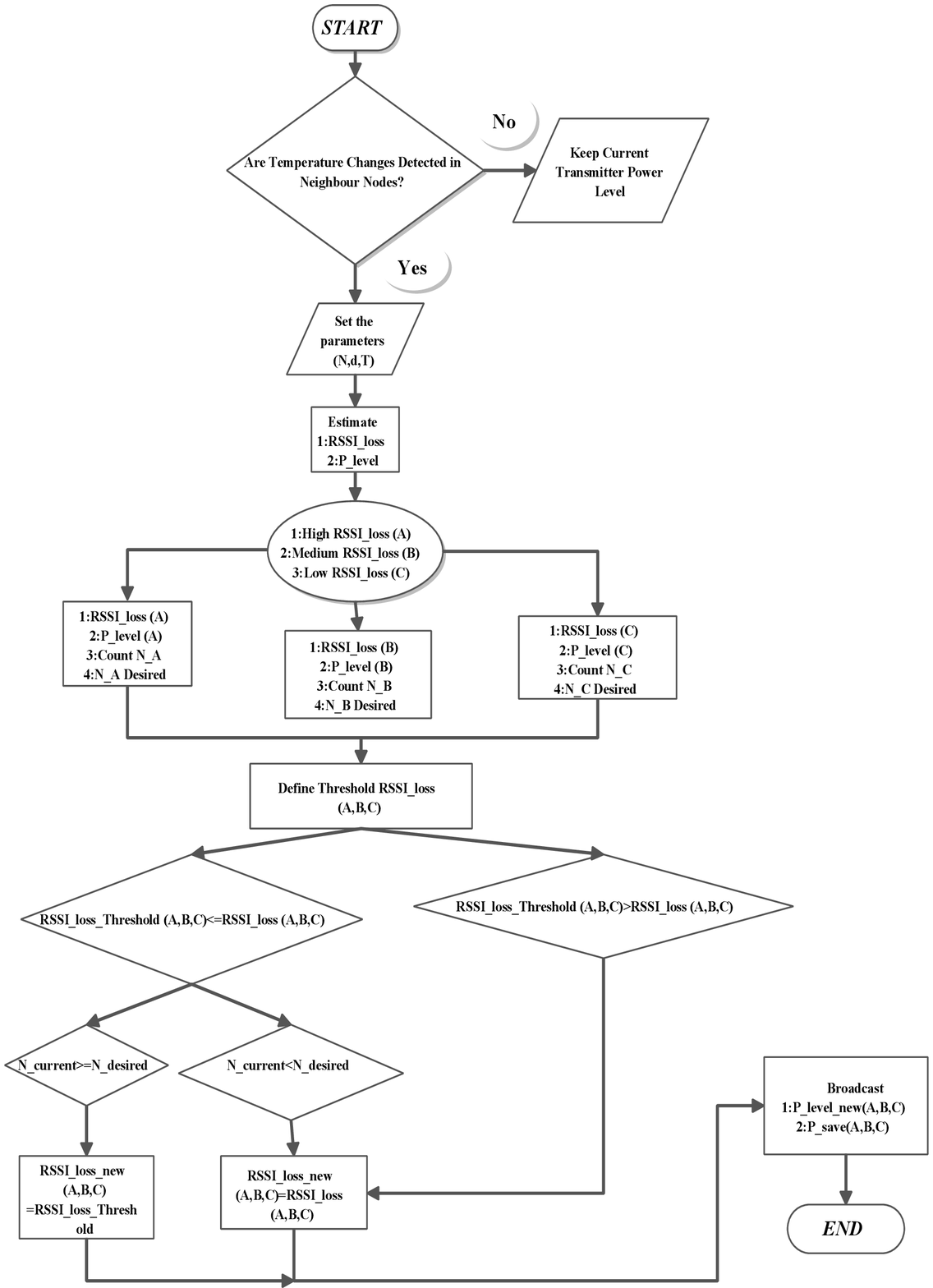}
\caption{Flow chart of Reference Node}
\end{center}
\end{figure*}

Fig1 shows system block diagram of proposed scheme. PRR, ACK, and $RSSI_{loss}$ used to determine connectivity. ACK estimates connectivity but it cannot determine link quality. PRR estimates connectivity accurately but it causes significant overhead ~\cite{8}. In our scheme, we use $RSSI_{loss}$ for connectivity estimation, which measures connectivity with relatively low overhead.

Power controller adjusts transmission $P_{level}$ by utilizing both number of current nodes and temperature sensed at each node. Since power controller is operated not merely by comparing number of current nodes with desired nodes but by using temperature-compensated $P_{level}$, so that it can reach to desired $P_{level}$ rapidly. If temperature is changing then temperature compensation is executed on basis of relationship between temperature and $RSSI_{loss}$. Network connectivity maintained with low overhead by reducing feedback process between nodes which is achieved due to logical division of network.

Transmission power loss due to temperature variation formulated using relationship between $RSSI_{loss}$ and temperature experimented in Bannister et al.. Mathematical expression for $RSSI_{loss}$ due to temperature variation is as follows ~\cite{12}:

\begin{equation}
RSSI_{loss}[dBm]=0.1996*(T[C^{o}]-25[C^{o}])
\end{equation}

To compensate $RSSI_{loss}$ estimated from Eq.(1) we have to control output $P_{level}$ of radio transmitter accordingly. Relationship between required transmitter $P_{level}$ and $RSSI_{loss}$ is formulated by Eq.(2) using least square approximation ~\cite{12}:

\begin{equation}
P_{level}=[(RSSI_{loss}+40)/12]^{2.91}
\end{equation}

Based on Eqs (1, 2), we obtain appropriate $P_{level}$ to compensate $RSSI_{loss}$ due to temperature variation. To compensate path loss due to distance  between each sensor node in WSN, free space model helps to estimate actual required transmitter power. After addition of $RSSI_{loss}$ due to temperature variation in Eq.(3), we estimate actual required transmitter power between each sensor node. For free space path loss model  we need number of nodes in a network (N), distance between each node (d), ($E_{b}/N_{o}$) depends upon ($SNR$), spectral efficiency ($\eta$), frequency ($f$) and receiver noise figure ($(RNF)$):
 \begin{equation}
P_{t}[dBm]=[\eta*(E_{b}/N_{0})*mkTB*(4\pi d /\lambda )^2+RNF]+RSSI_{loss}
\end{equation}

Parameters for propose scheme are,(1) Threshold $RSSI_{loss}$  for each region. (2) Desired nodes in each region $n_{d}(t)=n_{c}(t)-5$, (3) Transmission power level $P_{level}$ for each region.

Threshold $RSSI_{loss}$ is minimum value required to maintain link reliability. Reference node broadcasts beacon message periodically to nodes and wait for ACKs. If ACKs are received from nodes then $RSSI_{loss}$ is estimated for logical division of network, number of nodes with high $RSSI_{loss}$  considered in region A, medium $RSSI_{loss}$ considered in region B, and with low $RSSI_{loss}$ in region C. If ($RSSI_{loss}$ $\geq$ $RSSI_{loss}$ Threshold) and  ($n_{c}(t)\geq  n_{d}(t)$) then Threshold transmitter $P_{level}$ assigned if  for similar case ($n_{c}(t)<  n_{d}(t)$) then similar transmitter $P_{level}$ assigned and if  ($RSSI_{loss}$ $<$ $RSSI_{loss}$  Threshold) then by default keep same transmitter $P_{level}$. Given below is an algorithm for EAST.
\begin{algorithm}
\caption{EAST Algorithm}
\begin{algorithmic}[1]
\STATE $r \gets$ $Number$ $of$ $rounds$
\STATE $N \gets$ $Number$ $of$ $nodes$ $in$ $Network$
\STATE $d \gets$ $Distance$ $between$ $each$  $node$ $and$ $reference$ $node$
\STATE $T \gets$ $Temperature$ $for$ $each$ $node$
\STATE $RSSI_{loss} \gets$ $Transmission$ $power$ $loss$ $for$ $each$ $node$
\STATE $P_{level} \gets$ $Power$ $level$ $for$ $each$ $node$
\STATE $P_{t} \gets$ $Transmitter$ $power$ $for$ $each$ $node$
\STATE $Region$ $A \gets$  $High RSSI_{loss}$
\STATE $Region$ $B\gets$ $Medium RSSI_{loss}$
\STATE $Region$ $C\gets$ $Low RSSI_{loss}$
\STATE $n_{c}(t) \gets$ $Current$ $number$ $of$ $nodes$
\STATE $n_{d}(t) \gets$ $Desired$ $number$ $of$ $nodes$
\IF{$RSSI_{loss}(A,B,C)\geq RSSI_{loss}(Threshold)$}
\IF{$n_{c}(t)(A,B,C)\geq n_{d}(t)(A,B,C)$}
\STATE $RSSI_{loss}(new)(A,B,C)=RSSI_{loss}(Threshold)$
\ELSE
\STATE $RSSI_{loss}(new)(A,B,C)=RSSI_{loss}(A,B,C)$
\ENDIF
\ENDIF
\IF{$RSSI_{loss}(A,B,C)< RSSI_{loss}(Threshold)$}
\STATE $RSSI_{loss}(new)(A,B,C)=RSSI_{loss}(A,B,C)$
\ENDIF
\STATE $P_{levsl}(Save)(A,B,C) = P_{level}- P_{level}(new)(A,B,C)$
\end{algorithmic}
\end{algorithm}

Fig2 shows complete flow chart for reference node. Node senses temperature by using locally installed sensor and checks if temperature change detected. If there is any temperature change, compensation process is executed on the basis of Eqs (1, 2). Nodes send an ACK message including temperature change information with a newly calculated $P_{level}$. Apply ing this temperature-aware compensation scheme we can reduce overhead caused by conventional scheme in changing temperature environments.

\section{Mathematical Representation of the Proposed Scheme}
Let suppose we have 100 nodes in a network that are randomly deployed represented as ($N_{i}$). Nodes are  placed at different locations in a square area of 100*100m and distance ($d_{i}$) between them is from 1 to 100m. For given environment temperature ($T_{i}$) can have values  in range
-10$C^{0}$ $\leq$ $T_{i}$ $\leq$ 53 $C^{0}$   $\forall$ i $\epsilon$ N.\\
$RSSI_{loss}$ due to the temperature variation can be formulated using the relation between $RSSI_{loss}$ and the temperature experimented in Bannister et al ~\cite{12}. Equation for the $RSSI_{loss}$ for the temperature variation is as follows:\\

\begin{equation}
RSSI_{loss}(i)[dBm]=0.1996*(T_{i}[C^{o}]-25[C^{o}])
\end{equation}
Relation between$P_{level}$ and $RSSI_{loss}$ is formulated by using a least square approximation~\cite{12}:\\
\begin{equation}
P_{level}(i)=[(RSSI_{loss}(i)+40)/12]^{2.91}
\end{equation}
Maximum, minimum and average value of $RSSI_{loss}$  for all nodes in network can be formulated as:\\
\begin{equation}
RSSI_{loss}(min)= min(RSSI_{loss}(i))
\end{equation}

\begin{equation}
RSSI_{loss}(max)= max(RSSI_{loss}(i))
\end{equation}

\begin{equation}
RSSI_{loss}(avg)= (min(RSSI_{loss}(i))+max(RSSI_{loss}(i)))/2
\end{equation}
After finding maximum and minimum values of $RSSI_{loss}$ we will define upper and lower limit of $RSSI_{loss}$ to divide network into three regions and also set counter to count number of nodes in each region. Let suppose we have set counter zero initially and then define upper and lower bound and check condition, nodes that follow this condition are considered to be in region A $\forall$ i $\epsilon$ N.\\
\begin{equation}
RSSI_{loss}(Amax)= max(RSSI_{loss}(i))
\end{equation}
\begin{equation}
RSSI_{loss}(Amin)=RSSI_{loss}(avg)+2
\end{equation}
count=0;\\
$count_{A}$=count+1\\
Given that $\forall$ i $\epsilon$ N;\\
$RSSI_{loss}(i)$ $\leq$ $RSSI_{loss}(Amax)$ and $RSSI_{loss}(i)$ $>$ $RSSI_{loss}(Amin)$\\
Similarly we define upper and lower limits for region B and C and also check nodes that follow given conditions are said to be in region B and C respectively.\\
\begin{equation}
RSSI_{loss}(Bmax)=RSSI_{loss}(avg)+2
\end{equation}
\begin{equation}
RSSI_{loss}(Bmin)=RSSI_{loss}(avg)-2
\end{equation}
count=0;\\
$count_{B}$=count+1\\
Given that $\forall$ i $\epsilon$ N;\\
$RSSI_{loss}(i)$ $\leq$ $RSSI_{loss}(Bmax)$ and $RSSI_{loss}(i)$ $>$ $RSSI_{loss}(Bmin)$\\
\begin{equation}
RSSI_{loss}(Cmin)= min(RSSI_{loss}(i))
\end{equation}
\begin{equation}
RSSI_{loss}(Cmax)=RSSI_{loss}(avg)-2
\end{equation}
count=0;\\
$count_{C}$=count+1\\
Given that $\forall$ i $\epsilon$ N;\\
$RSSI_{loss}(i)$ $\leq$ $RSSI_{loss}(Cmax)$ and $RSSI_{loss}(i)$ $\geq$ $RSSI_{loss}(Cmin)$\\
To apply our proposed scheme $EAST$ we need to define threshold on $RSSI_{loss}$ for each region for energy efficient communication between sensor nodes. Threshold on $RSSI_{loss}$ for each region depends upon $RSSI_{loss}$ of all nodes in a particular region and number of nodes in that region. Threshold on $RSSI_{loss}$ for each region is defined as:\\
\begin{equation}
RSSI_{loss}(Threshold_{A})=\sum_{i=1}^{count_{A}} (RSSI_{loss}(i))/count_{A}
\end{equation}
\begin{equation}
RSSI_{loss}(Threshold_{B})=\sum_{i=1}^{count_{B}} (RSSI_{loss}(i))/count_{B}
\end{equation}
\begin{equation}
RSSI_{loss}(Threshold_{C})=\sum_{i=1}^{count_{C}} (RSSI_{loss}(i))/count_{C}
\end{equation}
$PRR$ is also an important metric to measure link reliability. Here $count_{A}$ are $n_{d}(t)$ and $count_{\bar {A}}$ number of nodes not present in region due to mobility and ($count_{A}$-$count_{\bar {A}}$) are $n_{c}(t)$. It is defined as number of nodes present in a region at particular time $n_{c}(t)$ to number of desired nodes $n_{d}(t)$ in  a region.  Similarly we can define $PRR$ for regions B and C. $PRR$ for all three regions is defined as given below:\\
\begin{equation}
PRR_{A}=(count_{A}-count_{\bar {A}})/count_{A}
\end{equation}
\begin{equation}
PRR_{B}=(count_{B}-count_{\bar {B}})/count_{B}
\end{equation}
\begin{equation}
PRR_{C}=(count_{C}-count_{\bar {C}})/count_{C}
\end{equation}
Here $PRR_{A}$, $PRR_{B}$ and $PRR_{C}$  are packet reception ratio for regions A, B, C respectively. $RSSI_{loss}$ for each region on basis of propose scheme for given conditions like threshold $RSSI_{loss}$ and $n_{c}(t)$ is formulated as:  \\
\begin{equation}
RSSI_{loss}(\tilde {A}, \tilde {B}, \tilde {C})(i)=RSSI_{loss} (Threshold A, B, C)
\end{equation}
Given that $\forall$ i $\epsilon$ N:\\
$RSSI_{loss}(Threshold A, B, C)$ $\leq$ $RSSI_{loss}(A, B, C)(i)$ and $n_{c}(t)(A, B, C)$ $\geq$ $n_{d}(t)(A, B, C)$\\
\begin{equation}
RSSI_{loss}(\tilde {A}, \tilde {B}, \tilde {C})(i)=RSSI_{loss}(A, B, C)(i)
\end{equation}
Given that $\forall$ i $\epsilon$ N:\\
$RSSI_{loss}(Threshold A, B, C)$ $\leq$ $RSSI_{loss}(A, B, C)(i)$ and $n_{c}(t)(A, B, C)$ $\leq$ $n_{d}(t)(A, B, C)$ or $RSSI_{loss}(Threshold A, B, C)$ $>$ $RSSI_{loss}(A, B, C)(i)$\\
Estimation of $P_{level}$ for new $RSSI_{loss}$ is formulated as $\forall$ i $\epsilon$ N:\\
\begin{equation}
P_{level}(\tilde {A}, \tilde {B}, \tilde {C})(i)=[(RSSI_{loss}(\tilde {A}, \tilde {B}, \tilde {C})(i)+40)/12]^{2.91}
\end{equation}
$P_{save}$ is defined as the difference between $P_{level}$ assigned before applying propose scheme and after applying propose scheme:\\
\begin{equation}
P_{save}(A, B, C)=\sum_{i=1}^{N} (P_{level}(A, B, C)(i))- \sum_{i=1}^{N} (P_{level}(\tilde {A}, \tilde {B}, \tilde {C})(i))
\end{equation}
Network life time can be enhanced by maximizing $P_{save}$. Aim of proposed scheme is to save maximum power with link reliability. Objective  function formulation for $P_{save}$ is defined $\forall$ i $\epsilon$ N:\\
\begin{equation}
Maximize   \sum_{i=1}^{N} (P_{save}(i))
\end{equation}
Constraints to save maximum power are given below $\forall$ i $\epsilon$ N:\\
\begin{equation}
  \sum_{i=1}^{N} RSSI_{loss}(A, B, C)(i)\geq  RSSI_{loss}(Threshold A, B, C)
\end{equation}
\begin{equation}
  \sum_{i=1}^{N} n_{c}(t)(A, B, C)(i)\geq \sum_{i=1}^{N} n_{d}(t)(A, B, C)(i)
\end{equation}
\begin{equation}
  \sum_{i=1}^{N} count_{AT}(A, B, C)(i)\geq \sum_{i=1}^{N} count_{BT}(A, B, C)(i)
\end{equation}
Here $count_{AT}$ and $count_{BT}$ are number of nodes above and below threshold in each region respectively.

\section{Results and Discussions}

In this section, we describe simulation results of proposed technique for energy efficient transmission in WSNs. Simulation parameters are; rounds 1200, temperature -10-53 $C^{0}$, distance (1-100)m, nodes 100, regions A,B,C, $\eta$ 0.0029, SNR 0.20dB, bandwidth 83.5MHz, frequency 2.45GHz, RNF 5dB, T 300k, $E_{b}/N_{0}$  8.3dB.  In Fig3 we have shown values of meteorological temperature for one round that each sensor node have sensed. Let suppose we have 100 nodes in 100*100 $m^{2}$ square region and temperature can have values in range (-10 - 53)$C^{o}$  ~\cite{13} for given meteorological condition of Pakistan. Reference node is placed at edge of this region.

%fig3
 \begin{figure}[h]
\begin{center}
\includegraphics[scale=0.32]{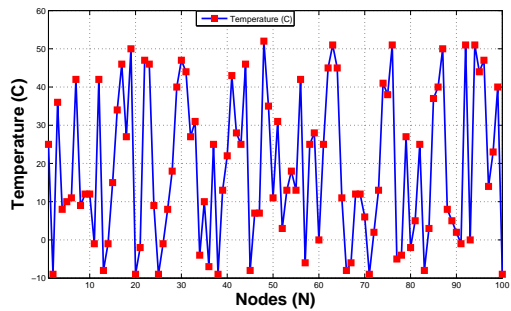}
\caption{Temperature sensed at each sensor node}
\end{center}
\end{figure}
%fig4
 \begin{figure}[h]
\begin{center}
\includegraphics[scale=0.32]{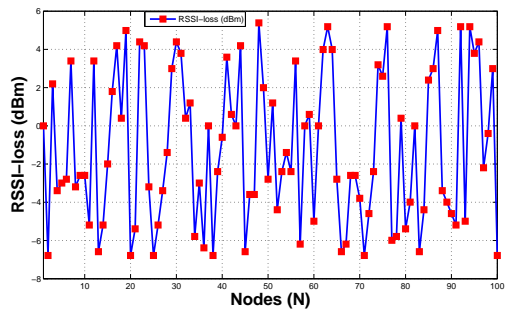}
\caption{Estimated Transmission power loss}
\end{center}
\end{figure}

%fig5
 \begin{figure}[h]
\begin{center}
\includegraphics[scale=0.32]{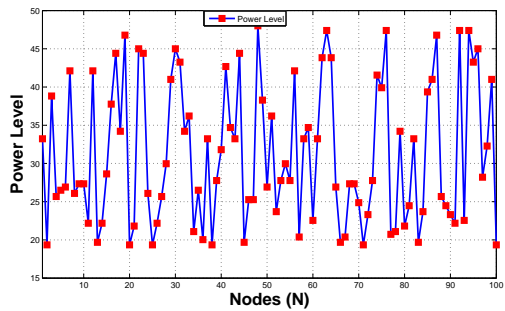}
\caption{Required Power level}
\end{center}
\end{figure}

%fig6
 \begin{figure}[h]
\begin{center}
\includegraphics[scale=0.32]{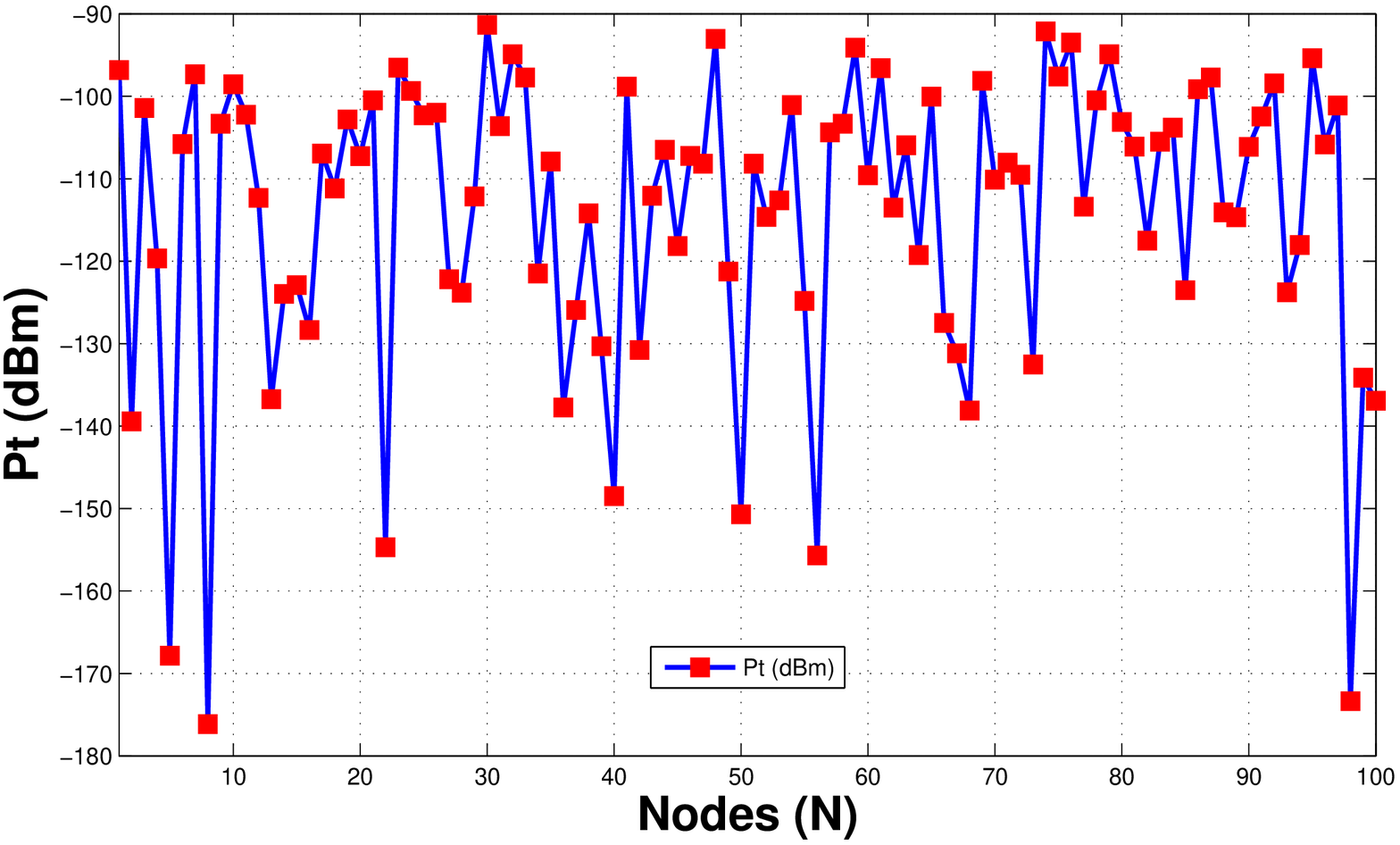}
\caption{Transmitter Power}
\end{center}
\end{figure}

Different values of temperature for each sensor node based on meteorological condition  helps to estimate $RSSI_{loss}(dBm)$. Fig4 shows $RSSI_{loss}(dBm)$ due to temperature variation in any environment using the relationship between $RSSI_{loss}(dBm)$ and temperature $(C^{o})$ given by Bannister et al.  High $RSSI_{loss}(dBm)$  means that sensor node  placed in region where temperature is high so link not have good quality. For temperature (-10 - 53)$C^{o}$  $RSSI_{loss}(dBm)$ have value in range (-6dBm) - (5dBm).

\begin{table}[h!]
  \centering
  \caption{Estimated Parameters}
  \tiny
  \begin{tabular}{|c|c|}\hline
    N (A,B,C)   & 46,30,24 \\ \hline
    $n_{d}$  (A,B,C)   & 41,25,19 \\ \hline
    $n_{c}$ (A,B,C) & 41,22,17 \\ \hline
    Threshold power level (A,B,C)  & 43.24,31.77,22.21 \\ \hline
    Nodes above threshold $RSSI_{loss}$ (A,B,C)  & 23,11,8 \\ \hline
    Nodes below threshold $RSSI_{loss}$ (A,B,C)   & 18,11,9 \\ \hline
    PRR (A,B,C) & (80-98),(70-96),(63-97) $\%$ \\ \hline
    Threshold $RSSI_{loss}$ ( A,B,C) & 3.78,-0.61,-5.17 dBm \\\hline
    \end{tabular}%
  \label{tab:addlabel}%
\end{table}%

From Fig4 it is also clear that link quality and $RSSI_{loss}$ have inverse relation, when temperature is high $RSSI_{loss}$ has high value means low quality link and vise versa. After estimating $RSSI_{loss}$ for each node in WSN we compute corresponding transmitter $P_{level}$ to compensate $RSSI_{loss}$.  Fig5 shows range of $P_{level}$ on y-axis for given $RSSI_{loss}$ that is between (20- 47) and also variation of required $P_{level}$ for sensor node with changing temperature that is at low temperature required $P_{level}$ is low and for high temperature required $P_{level}$ is high.

%fig7
 \begin{figure}[h]
\begin{center}
\includegraphics[scale=0.32]{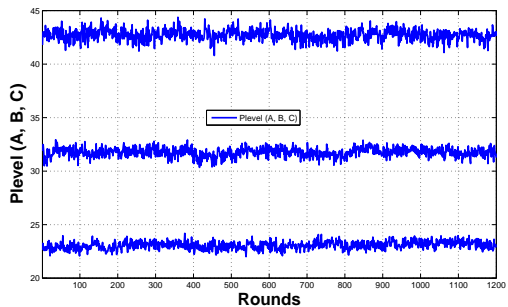}
\caption{Power level using Classical Approach for regions A, B, and C}
\end{center}
\end{figure}

As we have earlier estimated $RSSI_{loss}$ for each sensor node on the basis of given meteorological temperature that helps to estimate required $P_{level}$ to compensate $RSSI_{loss}$. That power level only helps to compensate $RSSI_{loss}$ due to temperature variations. To compensate path loss due to distance  between each sensor node in WSNs, free space model helps to estimate actual required transmitter power. After addition of required $P_{level}$ due to temperature variation and distance, we estimate actual required $P_{t}$ between each sensor node. Fig6 shows required $P_{t}$  including both $RSSI_{loss}$ due to temperature variation and free space path loss for different nodes. We clearly see from figure that $P_{t}$ lies between (-175 - 90)$dBm$ and most of times it is above -120$dBm$ .

%fig8
 \begin{figure}[h]
\begin{center}
\includegraphics[scale=0.32]{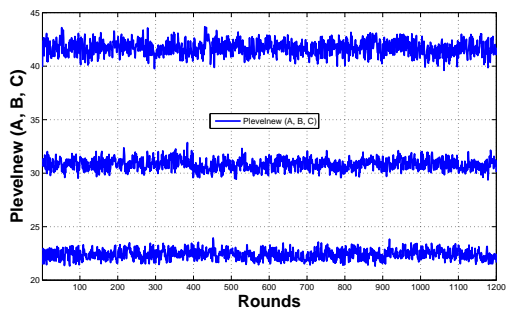}
\caption{ Power level using $EAST$ for regions A, B, and C}
\end{center}
\end{figure}
%fig9
 \begin{figure}[h]
\begin{center}
\includegraphics[scale=0.32]{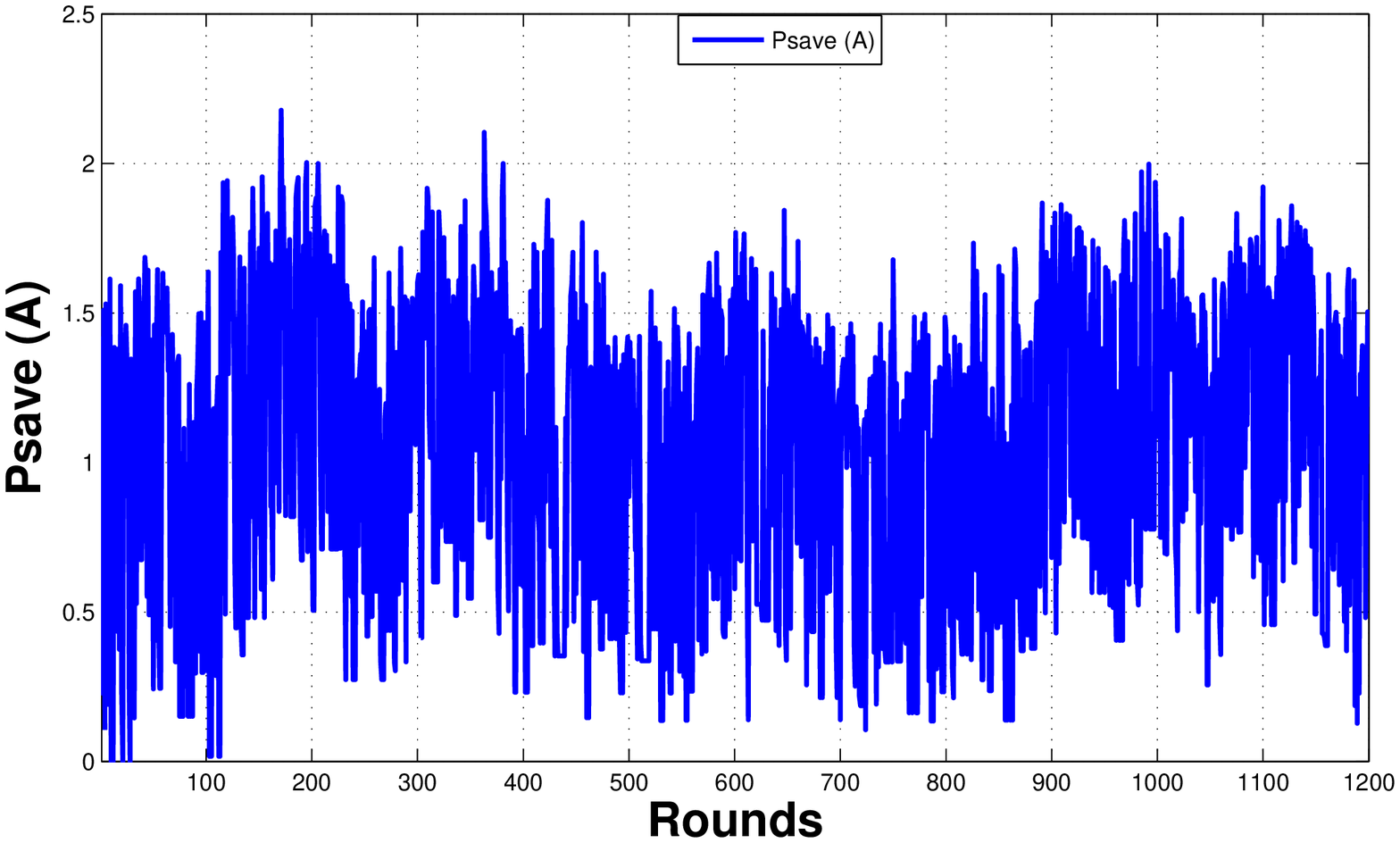}
\caption{Difference of Power level save between Classical Technique and EAST for region A}
\end{center}
\end{figure}
%fig10
 \begin{figure}[h]
\begin{center}
\includegraphics[scale=0.32]{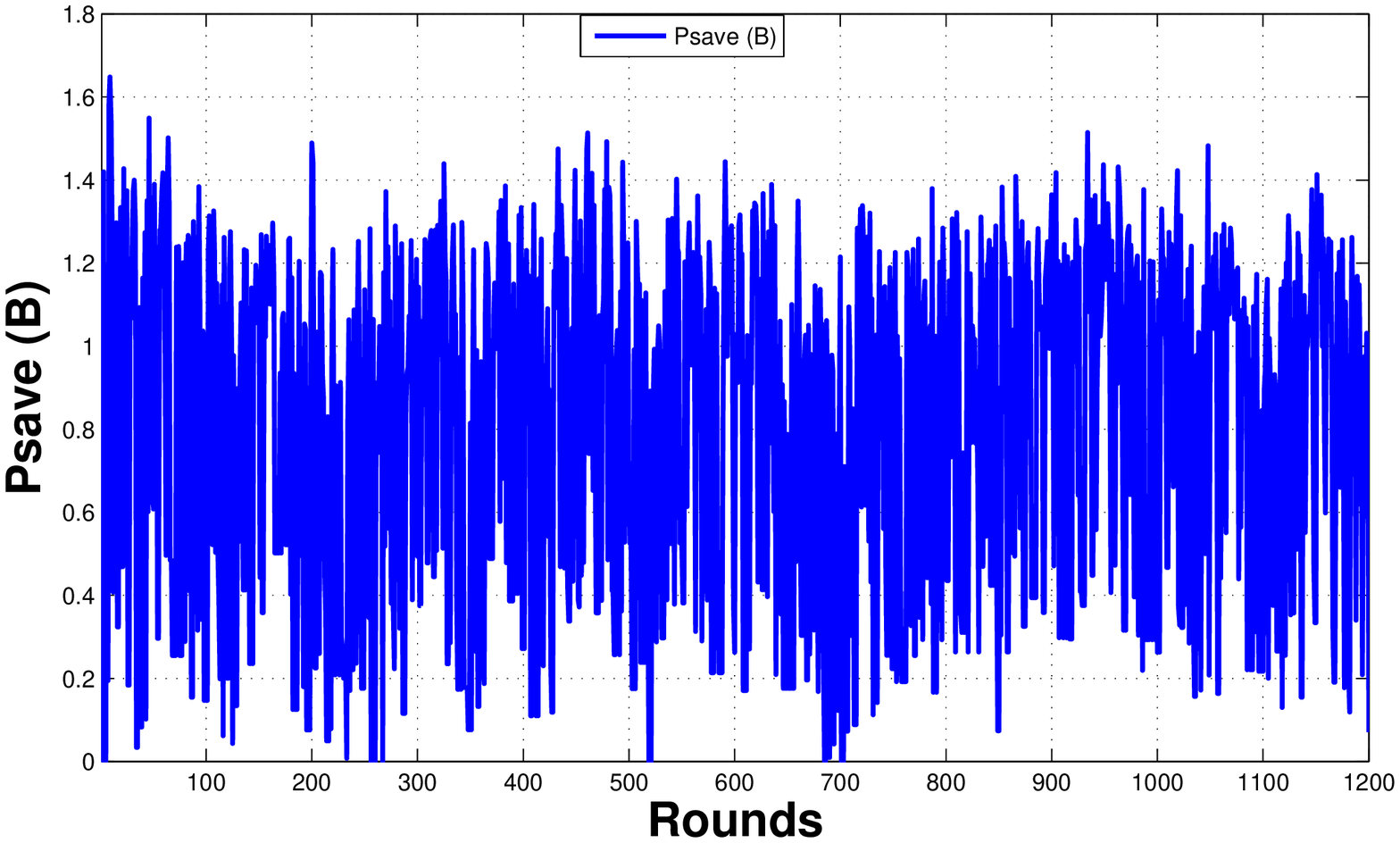}
\caption{Difference of Power level save between Classical Technique and EAST for region B}
\end{center}
\end{figure}
%fig11
 \begin{figure}[h]
\begin{center}
\includegraphics[scale=0.32]{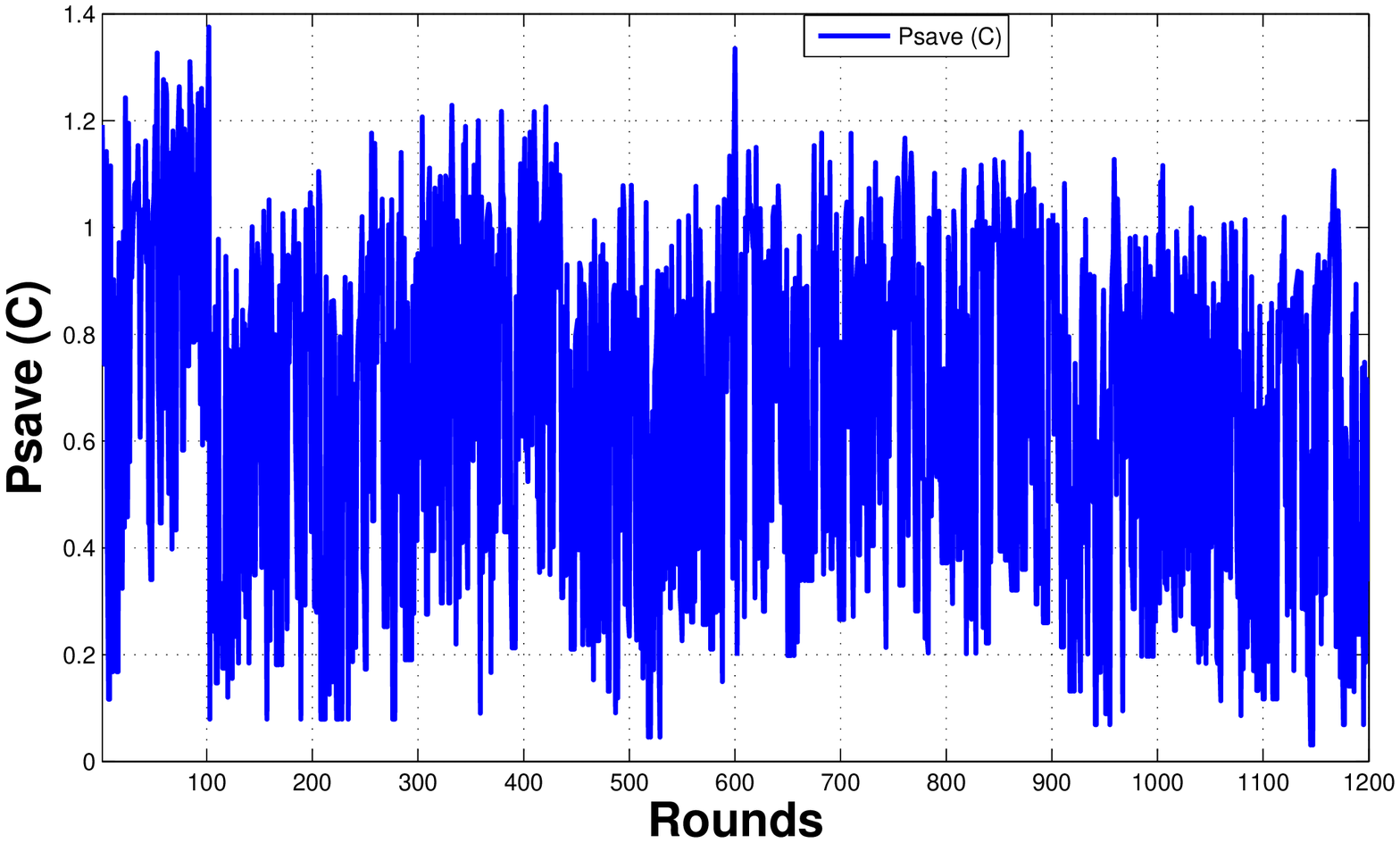}
\caption{Difference of Power level save between Classical Technique and EAST for region C}
\end{center}
\end{figure}

In Fig7, we have shown $P_{level}$ using classical approach for three regions and in Fig8, $P_{level}$ for the proposed technique; EAST. We can clearly see the difference between $P_{level}$ assigned. To show $P_{level}$ for each region, we take the difference between the assigned $P_{level}$s using EAST and classical technique, as can be seen in the figures 9, 10, 11. As we know that in classical approach, there is no concept of sub regions, so, for the sake of comparison with the proposed technique; EAST, we have shown $P_{level }$ for different regions using classical approach.

After estimating $RSSI_{loss}$ for nodes of each region, we have estimated required $P_{level}$ for nodes of each region that we clearly see in Fig7, in region A, $P_{level}$ lies between (40-45), for region B (30-35) and for region C (20-25). It means that for region A required $P_{level}$ high than both other region that also shows that for that region temperature and $RSSI_{loss}$ is large. For region B required $P_{level}$ is between both region A and C and for C region required $P_{level}$ is less than both other two regions. We have earlier seen in Fig7 $P_{level}$ for each region assigned using classical approach. After applying proposed technique we see what $P_{level}$ required for each region. We can clearly see difference  between $P_{level}$ as shown in Fig8, that required $P_{level}$ decreases for each region and for region A it decreases maximum. Fig9,10,11 respectively  shows required $P_{save}$ for region A,B and C after implanting proposed technique. $P_{save}$ up to 2.3 for region A, 1.7 for B and 1.5 for C.

Fig12 describes the effect of reference node mobility on $P_{save}$ for region A.   Reference node move around boundaries of  square region and nodes in a region  considered to be static. When reference node is at center location (50, 50) of network maximum nodes around reference node have large $RSSI_{loss}$ than threshold so we need to reduce $P_{level}$ to meet threshold $P_{level}$ requirement that cause maximum $P_{save}$.  We can clearly see maximum $P_{save}$ 12dBm to 20dBm for center location.  When reference  node move  from center to one of  the corner (0, 0) of square region $P_{save}$ remains constant approximately around 1dB, fact is that number of nodes near reference node region having same $RSSI_{loss}$ mean constant temperature and they need approximately same $P_{level}$ near threshold. $P_{save}$ for reference node movement from (0, 0) to (0, 100) fluctuate between -5dBm - 6dBm and at two moments we observe maximum $P_{save}$ because number of nodes near reference node have to increase their $P_{level}$ to meet threshold is minimum.

Movement of reference node from (0, 100) to (100, 100) causes $P_{save}$ between -4dBm - 12dBm and only one time peak $P_{save}$. Similarly when reference node move from (100, 100) to (100, 0) $P_{save}$ remains in limits between -4dBm- 7dBm and only one time maximum $P_{save}$ . From this figure it is also clear that for region A reference node location at center gives maximum $P_{save}$ that enhances network lifetime. We can also see variation of $P_{save}$ with respect to time that basically depends upon nodes near reference node have what $RSSI_{loss}$ if nodes have less $RSSI_{loss}$ then threshold then we have to increase $P_{level}$ that decrease $P_{save}$ and if nodes have large $RSSI_{loss}$ then threshold then we need to decrease $P_{level}$ that enhances $P_{save}$. It is also clear from result that peak maximum and minimum $P_{save}$ comes at same time.

%fig12
 \begin{figure}[h]
\begin{center}
\includegraphics[scale=0.32]{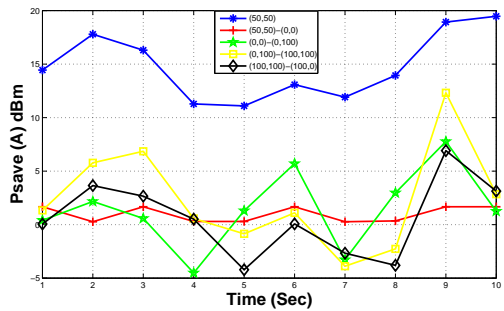}
\caption{ Transmitter power save in region A for different Reference Node Locations}
\end{center}
\end{figure}

Similarly we can see $P_{save}$ for similar pattern of reference node mobility considering regions B and C. For region B in Fig13 when reference node at center location (50, 50) $P_{save}$  remains between 14dBm-20dBm, from center to (0, 0) $P_{save}$ remains between 0 - 1dBm. When reference node moves from (0, 0) to one of the corner of square region (0, 100) $P_{save}$  fluctuate between 0 - 4dBm. Reference node movement from (0, 100) to (100, 100)  cause $P_{save}$ -1dBm-5dBm. Reference node movement from (100, 100) to (100, 0) $P_{save}$ -4dBm-5dBm.

This figure also indicates  that $P_{save}$ for region B is maximum when reference node at center location. For reference node mobility from center to (0, 0) $P_{save}$ remains constant due to constant $RSSI_{loss}$ near reference node region. For other reference node movements $P_{save}$ remains approximately constant due to less variations in $RSSI_{loss}$. Compared to region A where $P_{save}$ goes to peak maximum and minimum value in region B $P_{save}$ remains on average approximately constant and less variation occurs,  fact is that nodes in region B have approximately same $RSSI_{loss}$ near threshold.

%fig13
 \begin{figure}[h]
\begin{center}
\includegraphics[scale=0.32]{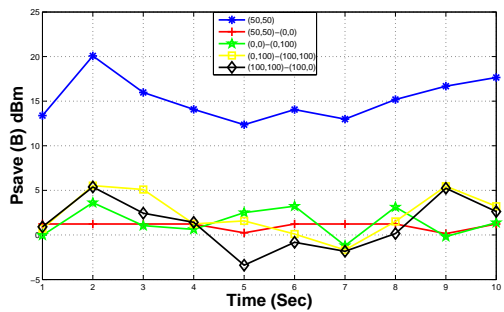}
\caption{ Transmitter Power save in region B for different Reference Node Locations}
\end{center}
\end{figure}

$P_{save}$ for reference node mobility in region C around square as shown in Fig14. When reference node is at center (50, 50)  $P_{save}$ fluctuates between 8dBm-50dBm. From center to edge (0, 0) reference node mobility cause $P_{save}$ around 0dBm. When reference node move from corner of square (0, 0) to corner (0, 100) $P_{save}$ -5dBm-12dBm. Similarly from (0, 100) to (100, 100) $P_{save}$ remains between -10dBm-18dBm.  Finally when reference node location changes from (100, 100) to (100, 0) $P_{save}$ goes to maximum value 60dBm that shows that nodes near reference node have large $RSSI_{loss}$ than threshold $RSSI_{loss}$ at that moment. This figure also elaborates that on average $P_{save}$ maximum for reference node location at center. Compared to region B in this region peak maximum and minimum $P_{save}$ exists reason is that nodes in this region have large $RSSI_{loss}$ than threshold at that moment.
%fig14
 \begin{figure}[h]
\begin{center}
\includegraphics[scale=0.32]{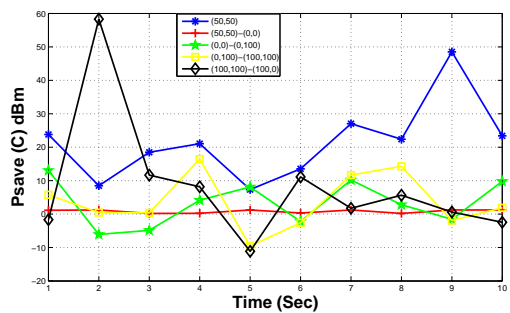}
\caption{ Transmitter Power save in region C for different Reference Node Locations}
\end{center}
\end{figure}

\section{Conclusion and Future Work}

In this paper, we presented a new proposed technique  EAST. It shows that temperature is one of most important factors impacting link quality. Relationship between $RSSI_{loss}$ and temperature has been analyzed for our transmission power control scheme. Proposed scheme uses open-loop control to compensate for changes of link quality according to temperature variation. By combining both open-loop temperature-aware compensation and close-loop feedback control, we can significantly reduce overhead of transmission power control in WSN, we  further extended our scheme by dividing network into three regions on basis of threshold $RSSI_{loss}$ and assign $P_{level}$ to each node in three regions on the basis of current number of nodes and desired number of nodes, which helps to adapt $P_{t}$ according to link quality variation and increase network lifetime. We have also evaluate the performance of propose scheme for reference node mobility around square region that shows $P_{save}$ up to 60dBm. But in case of static reference node $P_{save}$ goes maximum to 2dBm.

In future, firstly, we are interested to work on Internet Protocol (IP) based solutions in WSNs~\cite{14}. Secondly, as sensors are usually deployed in potentially adverse environments~\cite{15}, so, we will address the security challenges using the intrusion detection systems because they provide a necessary layer for the protection.

%\bibliography{mybibliography}
%\bibliographystyle{ieeetr}

\end{document}